\documentclass[10pt,letterpaper]{article}

\usepackage{opex3}
\usepackage{color}
\usepackage{threeparttable}
\usepackage[latin1]{inputenc} 
\usepackage{graphicx}        
\usepackage{subcaption}
\usepackage{amsmath}
\usepackage{amssymb}
\usepackage{textcomp}   

\newcommand{\ba}{\begin{eqnarray}}
\newcommand{\ea}{\end{eqnarray}}

\begin{document}

\title{Modeling Fluorescence Correlation Spectroscopy through an aberrating sphere}

\author{Daja Ruhlandt,$^{1}$ Aditya Katti,$^{1}$ Jacques Derouard,$^{2}$ \\Antoine Delon,$^{2}$
and J\"org Enderlein$^{1,3,*}$}

\address{$^{1}$III. Institute of Physics -- Biophysics, Georg August University, 37077 G\"ottingen, Germany\\
$^{2}$Universit\'e Grenoble Alpes, CNRS, LIPhy, 38000 Grenoble, France\\
$^{3}$Cluster of Excellence "Multiscale Bioimaging: from Molecular Machines to Networks of Excitable Cells" (MBExC), Georg August University, G\"ottingen, Germany}
\email{*jenderl@gwdg.de}
\homepage{www.joerg-enderlein.de}

\begin{abstract}
Fluorescence Correlation Spectroscopy (FCS) is a powerful single-molecule technique which allows for measuring motion (diffusion, flow), concentration, and molecular interaction kinetics of fluorescent molecules from picomolar to micromolar concentrations. It has found manifold applications in the physical and life sciences. Many biological/biophysical applications use FCS for measuring the motion and concentration of fluorescently labeled biomolecules in living cells and tissue. However, a correct quantitative evaluation of FCS experiments relies on the accurate knowledge of the fluorescence excitation and detection properties of the used confocal microscope. Using a bottom-up approach, we theoretically study how these properties are affected by the presence of a diffracting dielectric bead within the optical path, and how this changes the outcome of a FCS measurement. This will be important for all applications of FCS under optically non-ideal aberrating conditions. 
\end{abstract}



\section{Introduction}\label{sec:introduction}

Fluorescence Correlation Spectroscopy was invented by Magde, Elson and Webb in the early seventies of the last century~\cite{elson1974fluorescence,magde1974fluorescence}. It is based on confocal laser scanning microscope equipped with a high-sensitive detector and measures the temporal fluorescence intensity fluctuation which are excited in and detected from the tiny detection volume of the confocal microscope. On one hand, the temporal autocorrelation function of these intensity fluctuations delivers information about diffusion, directed motion~\cite{magde1978fluorescence}, intermolecular interaction~\cite{magde1972thermodynamic}, photo-physics~\cite{widengren2000characterization} or any other processes that affect the temporal evolution of the recorded fluorescence signal. On the other hand, the amplitude of the autocorrelation function is inversely proportional to the concentration of fluorescent molecules.

Mathematically, the autocorrelation function describes the intrinsic correlation of the recorded signal and measures the conditional probability to detect a photon at lag time $t$ if there was a photon detection event at time zero. This autocorrelation typically shows a lag-time dependent part, which contains information about real physical correlations, and a constant offset which is generated by the lag-time independent chance to detect two physically uncorrelated photons. The time dependent part of the autocorrelation function $g(t)$ is given by the product of the probability density to find a fluorescent molecule at location $\mathbf{r}_1$, which is equal to the concentration $c$ (number of molecules per volume), times the probability to see a fluorescence photon from this location, determined by the Molecule Detection Function (MDF) $U(\mathbf{r}_1)$ (probability density to excite \emph{and} detect a photon from a molecule located at position $\mathbf{r}$), times the probability that the molecule moves within time $t$ from position $\mathbf{r}_1$ to position $\mathbf{r}_2$ and is then still in a fluorescent state, described by a function $G(\mathbf{r}_2-\mathbf{r}_1,t)$, times the probability to detect a second photon from this new location, $U(\mathbf{r}_2)$, and the integration of this product over all possibilities, i.e. all initial and final positions $\mathbf{r}_1$ and $\mathbf{r}_2$. In mathematical terms, this reads

\begin{align}
\label{eq:autocorrelation}
g(t) = c \int d\mathbf{r}_2 \int d\mathbf{r}_1 U(\mathbf{r}_2) G(\mathbf{r}_2-\mathbf{r}_1,t) U(\mathbf{r}_1). 
\end{align}

\noindent For purely diffusing molecules and on time scale where all faster processes such as inter-system crossing to and return from a triplet state, or photo-isomerization, or other dark state kinetics have already decayed, the probability function $G(\mathbf{r}_2-\mathbf{r}_1,t)$ is the fundamental solution (Green's function) of the diffusion equation for the appropriate boundary conditions of the experiment (for example, open space for measurements in free solution, or reflecting boundary if the detection volume is crossed by an interface such as the glass surface of a cover slide or of a bead). The MDF $U(\mathbf{r})$ is given by the product of the excitation intensity distribution of the focused excitation laser times the detection efficiency distribution of the confocal detection.  In the next section, we consider first how to calculate the excitation intensity distribution when the laser is focused through a diffracting dielectric (glass) bead, and the section after the next, we will use this result for estimating the full MDF $U(\mathbf{r})$ and to calculate then the autocorrelation function $g(t)$. 

\section{Theory}\label{sec:basis}
\subsection{Focusing a laser beam through a spherical bead}

We start by considering the focusing of a linearly polarized laser beam through an ideal lens. Up to an overall constant, the electric field around the focal region (focus position $\mathbf{r}_0$) is described by the following plane wave superposition \cite{wolf1959electromagnetic,richards1959electromagnetic}:

\begin{equation}
\label{eq:efocus}
\begin{split}
\mathbf{E}_{f}(\mathbf{r}) = \iint_\Omega d\Omega \sqrt{\cos\chi} \; A(\chi) \left( \hat{\mathbf{e}}_p \cos\psi - \hat{\mathbf{e}}_s \sin\psi \right) \exp\left[ i k \hat{\mathbf{s}} \cdot \left( \mathbf{r} - \mathbf{r}_0 \right)  \right],
\end{split}
\end{equation}

\noindent where $A(\chi)$ is an amplitude and phase function (equal to some constant for perfect diffraction-limited focusing), $k$ is the length of the wave vector of the excitation light in the sample, the unit vectors $\hat{\mathbf{e}}_p$ and $\hat{\mathbf{e}}_s$ have, in Cartesian coordinates $\{x,y,z\}$, the components $\hat{\mathbf{e}}_p = \left\{ \cos\chi \cos\psi, \cos\chi \sin\psi, -\sin\chi \right\}$ and $\hat{\mathbf{e}}_s = \left\{ -\sin\psi, \cos\psi, 0 \right\}$, and each plane wave in the above superposition travels along the direction of unit vector $\hat{\mathbf{s}} = \left\{ \sin\chi \cos\psi, \sin\chi \sin\psi, \cos\chi \right\}$ with a polarization vector $\hat{\mathbf{p}} \perp \hat{\mathbf{s}}$ which can be represented as $\hat{\mathbf{p}} = \hat{\mathbf{e}}_p \cos\psi - \hat{\mathbf{e}}_s \sin\psi$. The integration in eq.~\eqref{eq:efocus} extends over the whole solid angle subtended by the cone of light collection of the focusing lens. 

\begin{figure}
    \centering
    \includegraphics[width=0.6\textwidth]{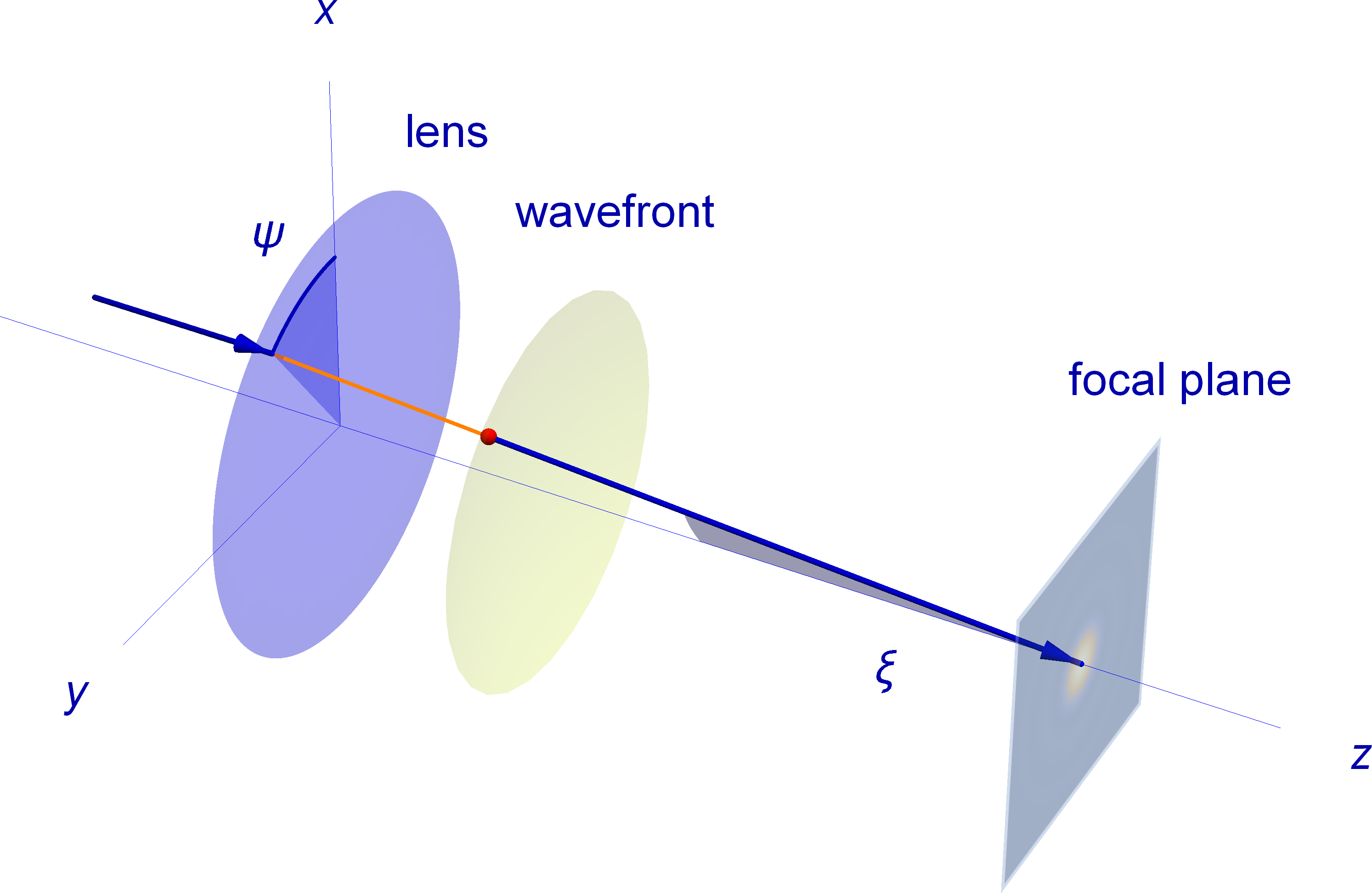}
    \caption{Schematic of focusing a planar wavefront through a lens.}
    \label{fig:focus}
\end{figure}

For modeling the interaction of such an electric field with a spherical bead of refractive index $n_g$ we will seek for an expansion of this electric field into the following basis functions which are based on vector spherical harmonics:

\begin{equation}
\begin{split}
\label{eq:M}
\mathbf{M}_{\ell m}^{(z)}\left(k \mathbf{r}\right) =  c_{\ell m} \left[\frac{i m P_{\ell}^m(\cos\theta)}{\sin\theta}\hat{\mathbf{e}}_\theta + \sin\theta {P_{\ell}^m}'(\cos\theta) \hat{\mathbf{e}}_\phi\right] z_\ell(k r) e^{i m \phi} 
\end{split}
\end{equation}

\noindent and 

\begin{equation}
\begin{split}
\label{eq:N}
\mathbf{N}_{\ell m}^{(z)}\left(k \mathbf{r}\right) = c_{\ell m} \Biggl\{ \ell (\ell +1) &P_{\ell}^m(\cos\theta) z_\ell(k r) \hat{\mathbf{e}}_r - \\
&-\frac{\partial(r z_\ell(k r))}{\partial r} \left[\sin \theta {P_{\ell}^m}'(\cos\theta) \hat{\mathbf{e}}_\theta - \frac{i m P_{\ell}^m(\cos\theta)}{\sin\theta} \hat{\mathbf{e}}_\phi \right] \Biggr\} \frac{e^{i m \phi}}{k r},
\end{split}
\end{equation}

\noindent where the normalizing coefficients $c_{\ell m}$ are

\begin{equation}
\begin{split}
c_{\ell m} = \sqrt{\frac{2\ell+1}{4\pi}\frac{(\ell-m)!}{(\ell+m)!}}
\end{split}
\end{equation}

\noindent and the functions $z_\ell = z_\ell(k r)$ are either spherical Bessel functions $j_\ell$, or one of the two spherical Hankel functions $h^1_\ell$ or $h^2_\ell$. The superscript $(z)$ on $\mathbf{M}_{\ell m}^{(z)}$ and $\mathbf{N}_{\ell m}^{(z)}$ refers to these functions. The symbol $k = n k_0 = 2 \pi n/\lambda$ denotes the length of the wave vector in a medium with refractive index $n$ for light with vacuum wavelength $\lambda$, $P_\ell^m = P_\ell^m(\cos\theta)$ are associated Legendre functions, $\hat{\mathbf{e}}_{r,\theta,\phi}$ are unit vectors along the indicated directions, and a prime on $P_\ell^m$ denotes differentiation after its argument $\cos\theta$. These basis functions have the following normalization and orthogonality properties

\begin{align}
\begin{split}
\label{eq:MNnorm}
\left\langle \mathbf{M}_{\ell' m'}^{(z)*}(k \mathbf{r}) \cdot \mathbf{M}_{\ell m}^{(z)}(k \mathbf{r}) \right\rangle &= \delta_{\ell' \ell} \delta_{m' m} \ell(\ell+1) \left\vert z_\ell(k r)\right\vert^2 \\
\left\langle \mathbf{N}_{\ell' m'}^{(z)*}(k \mathbf{r}) \cdot \mathbf{N}_{\ell m}^{(z)}(k \mathbf{r}) \right\rangle &= \delta_{\ell' \ell} \delta_{m' m} \frac{\ell(\ell+1)}{k^2r^2} \left[ \ell(\ell+1) \left\vert z_\ell(k r)\right\vert^2 + \left\vert \frac{\partial (r z_\ell(k r))}{\partial r}\right\vert^2\right] \\
\left\langle \mathbf{M}_{\ell' m'}^{(z)*}(k \mathbf{r}) \cdot \mathbf{N}_{\ell m}^{(z)}(k \mathbf{r}) \right\rangle &= 0
\end{split}
\end{align}

\noindent where the triangular brackets are short-hand for the integration

\begin{align}
\left\langle \bullet \right\rangle \equiv \int_0^{2\pi} d\phi \int_0^\pi d\theta \sin\theta \; (\bullet)
\end{align}

\noindent Furthermore, these functions obey the symmetric differential equations

\begin{align}
\label{eq:sym}
\mathbf{rot}\; \mathbf{M}_{\ell m}^{(z)} = k \mathbf{N}_{\ell m}^{(z)}, \quad \mathbf{rot}\; \mathbf{N}_{\ell m}^{(z)} = k \mathbf{M}_{\ell m}^{(z)}
\end{align}

\noindent With these basis functions, the electric field $\mathbf{E}_D\left(\mathbf{r}\vert\mathbf{r}'\right)$ at position $\mathbf{r}$ of an electric dipole emitter at position $\mathbf{r}'$ takes the form (for $r<r'$):

\begin{align}
\label{eq:dipole}
\begin{split}
\mathbf{E}_D\left(\mathbf{r}\vert\mathbf{r}'\right)=\sum_{\ell=1}^\infty \frac{4\pi i k k_0^2}{\ell(\ell+1)} \sum_{m=-\ell}^\ell \left\{ \left[ \hat{\mathbf{p}}\cdot \mathbf{M}_{\ell m}^{(h^2)*}\left(k \mathbf{r}'\right) \right] \mathbf{M}_{\ell m}^{(j)}\left(k \mathbf{r}\right) + \left[ \hat{\mathbf{p}}\cdot \mathbf{N}_{\ell m}^{(h^2)*}\left(k \mathbf{r}'\right) \right] \mathbf{N}_{\ell m}^{(j)}\left(k \mathbf{r}\right) \right\}
\end{split}
\end{align}

\noindent where, here, Bessel functions of the first kind ($z\equiv j$) and Hankel functions of the second kind ($z\equiv h^2$) are used in the $\mathbf{M}_{\ell m}^{(z)}$ and $\mathbf{N}_{\ell m}^{(z)}$  of eqs.~\eqref{eq:M}	and \eqref{eq:N}. When shifting the dipole position towards infinity along the negative direction of some unit vector $\hat{\mathbf{s}}$ that obeys the condition  $\hat{\mathbf{p}}\cdot \hat{\mathbf{s}} = 0$, the field around the coordinate origin is approximated by the well-known far-field term of a dipole emitter

\begin{align}
\label{eq:dipolefar}
\begin{split}
\mathbf{E}_D\left(\mathbf{r}\vert\mathbf{r}'\right) \rightarrow k_0^2 \hat{\mathbf{p}}\frac{\exp\left(i k r'+i \hat{\mathbf{s}}\cdot\mathbf{r}\right)}{r'}
\end{split}
\end{align}

\noindent so that $r' e^{-i k r'} \mathbf{E}_D\left(\mathbf{r}\vert\mathbf{r}'\right)/k_0^2$ will approach the plane wave $\exp\left(i k \hat{\mathbf{s}}\cdot \mathbf{r}\right)$. The asymptotic behavior of $\mathbf{M}_{\ell m}^{(h^2)*}$ and $\mathbf{N}_{\ell m}^{(h^2)*}$ is given by

\begin{align}
\label{eq:asymptotic}
\begin{split}
&\mathbf{M}_{\ell m}^{(h^2)*}\left(k r',\theta',\phi'\right) \rightarrow \frac{c_{\ell m}}{i^{\ell+1}} \left[-\frac{i m P_\ell^m\left(\cos\theta'\right)}{\sin\theta'} \hat{\mathbf{e}}_{\theta'} + \sin\theta' {P_\ell^m}'\left(\cos\theta'\right) \hat{\mathbf{e}}_{\phi'} \right] \frac{e^{i k r' - i m \phi'}}{k r'} \\
&\mathbf{N}_{\ell m}^{(h^2)*}\left(k r',\theta',\phi'\right) \rightarrow -\frac{c_{\ell m}}{i^{\ell}} \left[\sin\theta' {P_\ell^m}'\left(\cos\theta'\right) \hat{\mathbf{e}}_{\theta'} + \frac{i m P_\ell^m\left(\cos\theta'\right)}{\sin\theta'} \hat{\mathbf{e}}_{\phi'} \right] \frac{e^{i k r' - i m \phi'}}{k r'}
\end{split}
\end{align}

\noindent Representing $\hat{\mathbf{s}}$ in Cartesian coordinates as $\left\{ \sin\chi \cos\psi, \sin\chi \sin\psi, \cos\chi \right\}$ and taking further into account that $\theta' = \pi-\chi$ and $\phi' = \pi +\psi$ as well as the symmetry of associated Legendre polynomials, $P_\ell^m(-x)=(-1)^{\ell+m}P_\ell^m(x)$, we obtain a vector spherical harmonics representation of a plane wave traveling along direction $\hat{\mathbf{s}}$ and having polarization vector $\hat{\mathbf{p}}$:

\begin{equation}
\label{eq:planewave}
\begin{split}
&\hat{\mathbf{p}}\exp\left[i k \hat{\mathbf{s}} \cdot \left( \mathbf{r} - \mathbf{r}_0 \right)\right] =\\
&\sum_{\ell=1}^\infty \sum_{m=-\ell}^\ell \frac{4\pi i^{\ell-1}}{\ell(\ell+1)}c_{\ell m} \exp\left[ -i k z_0 \cos\chi  -i k \rho_0 \sin\chi \cos\left(\psi - \phi_0\right) - i m \psi \right] \cdot \Biggl\{ \frac{m P_\ell^m(\cos\chi)}{\sin\chi} \\ 
&\left[ p_p \mathbf{M}_{\ell m}^j(k \mathbf{r}) - i p_s \mathbf{N}_{\ell m}^j(k \mathbf{r}) \right] + \sin\chi {P_\ell^m}'(\cos\chi) \left[i p_s \mathbf{M}_{\ell m}^j(k \mathbf{r}) - p_p \mathbf{N}_{\ell m}^j(k \mathbf{r}) \right]\Biggr\}
\end{split}
\end{equation}

\noindent where $\mathbf{r} = r \left\{ \sin\theta \cos\phi, \sin\theta \sin\phi, \cos\theta \right\}$, $\mathbf{r}_0 = \left\{ \rho_0 \cos\phi_0, \rho_0 \sin\phi_0, z_0 \right\}$, and we have used the unit vectors from eq.~\eqref{eq:efocus} so that $\hat{\mathbf{e}}_p \times \hat{\mathbf{e}}_s = \hat{\mathbf{s}}$ and $p_{p,s} = \hat{\mathbf{e}}_{p,s} \cdot \hat{\mathbf{p}}$.

As shown in eq.~\eqref{eq:efocus}, for a linearly polarized laser beam (polarization along $\psi=0$) focused by an ideal lens we have $p_p = \cos\psi$ and $p_s = -\sin\psi$, and we can analytically perform the integration over $\psi$ in eq.~\eqref{eq:efocus} by employing the Jacobi-Anger expansion

\begin{equation}
\label{eq:jacobianger}
\begin{split}
\exp\left[-i k \rho_0 \sin\chi \cos\left(\psi - \phi_0\right)\right] = \sum_{m=-\infty}^\infty i^{-m} J_m\left(k \rho_0 \sin\chi \right) \exp\left[ i m \left(\psi-\phi_0\right) \right]
\end{split}
\end{equation}

\noindent where the $J_m$ denote Bessel functions of the first kind. As a result we find 

\begin{equation}
\label{eq:efocuspsi}
\begin{split}
\mathbf{E}_{f}\left(\mathbf{r}\right) = \sum_{\ell=1}^\infty\sum_{m=-\ell}^\ell \left[ a_{\ell m}^{f} \mathbf{M}^{(j)}_{\ell m}\left(k \mathbf{r}\right) + b_{\ell m}^{f} \mathbf{N}^{(j)}_{\ell m}\left(k \mathbf{r}\right) \right]
\end{split}
\end{equation}

\noindent with the coefficients

\begin{equation}
\label{eq:coefa}
\begin{split}
a_{\ell m}^{f} = -\int_{0}^{\Theta} d\chi &\sin\chi \sqrt{\cos\chi} \; A(\chi) \frac{2\pi c_{\ell m} e^{-i k z_0 \cos\chi - i m \phi_0}}{i^{\ell +m}\ell(\ell+1)} \cdot \\
&\cdot \left[ J_{m-1}e^{i \phi_0} \left( \frac{m P_\ell^m}{\sin\chi} + \sin\chi {P_\ell^m}' \right) - J_{m+1}e^{-i \phi_0} \left( \frac{m P_\ell^m}{\sin\chi} - \sin\chi {P_\ell^m}' \right) \right]
\end{split}
\end{equation}

\noindent and

\begin{equation}
\label{eq:coefb}
\begin{split}
b_{\ell m}^{f} = -\int_{0}^{\Theta} d\chi &\sin\chi \sqrt{\cos\chi} \; A(\chi) \frac{2\pi c_{\ell m} e^{-i k z_0 \cos\chi - i m \phi_0}}{i^{\ell +m}\ell(\ell+1)} \cdot \\
&\cdot \left[ J_{m-1}e^{i \phi_0} \left( \frac{m P_\ell^m}{\sin\chi} + \sin\chi {P_\ell^m}' \right) + J_{m+1}e^{-i \phi_0} \left( \frac{m P_\ell^m}{\sin\chi} - \sin\chi {P_\ell^m}' \right) \right]
\end{split}
\end{equation}

\noindent while the argument of the Bessel functions is $k \rho_0 \sin\chi$. The remaining integration over $\chi$ in eqs.~\eqref{eq:coefa} and \eqref{eq:coefb} has to be done numerically. The corresponding expansion for the magnetic field is found from Faraday's law $\mathbf{B}=(i k_0)^{-1} \mathbf{rot}\,\mathbf{E}$ and using eqs.~\eqref{eq:sym}.

\begin{figure}
    \centering
    \includegraphics[width=\textwidth]{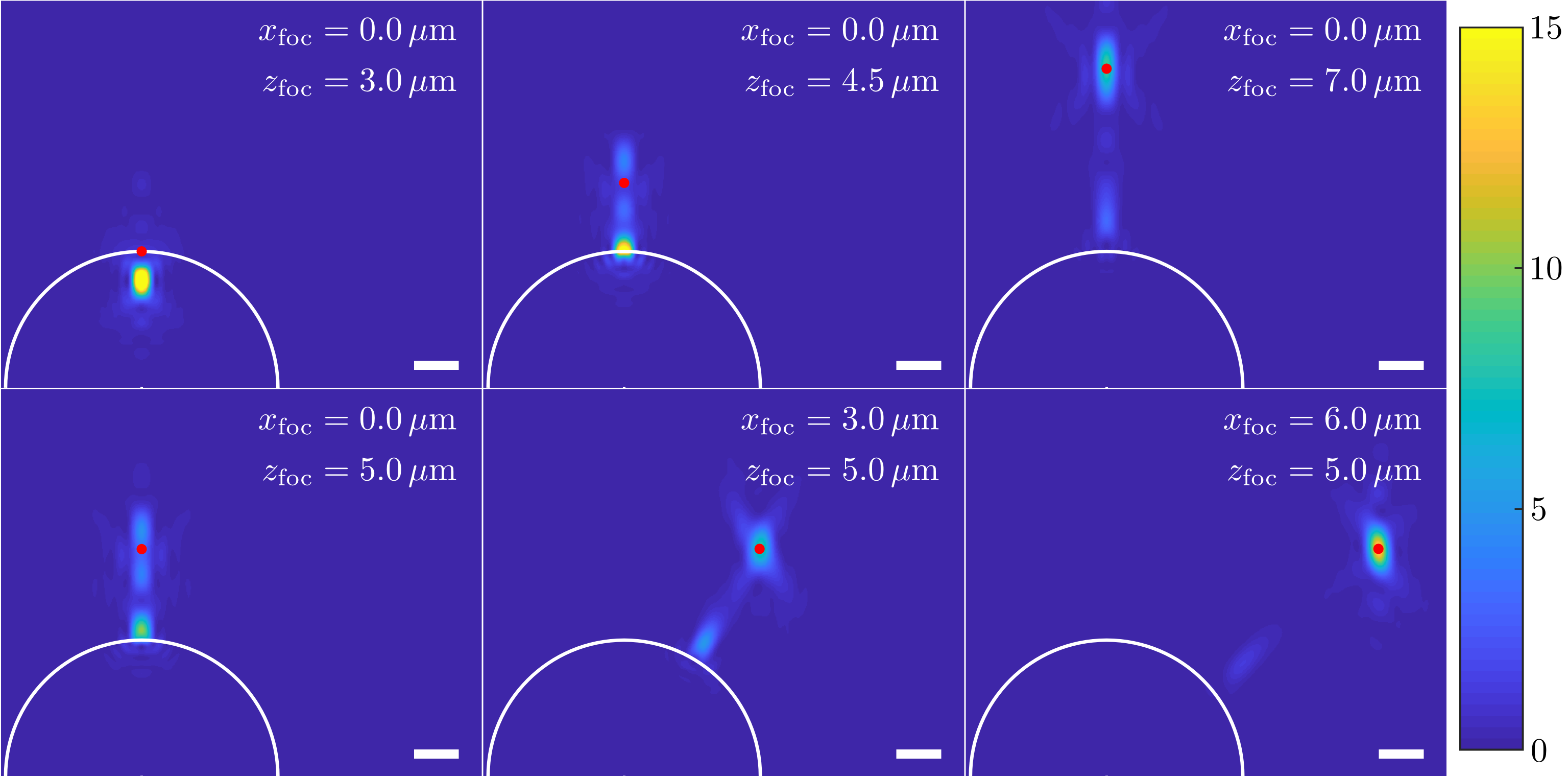}
    \caption{Plots of the excitation intensity distribution $|\mathbf{E}|^2$ of a plane wave focused in water close to a glass bead centered at the coordinate origin and with radius $R=3$\,\textmu m (contour shown as a white line). The images depict the $x$-$z$ plane for different nominal focus positions $\mathbf{r}_0=(x_{\rm{foc}},0,z_{\rm{foc}})$ as indicated by the red dots. Inside the bead, $\mathbf{E}=\mathbf{E}_i$ and outside the bead, $\mathbf{E}=\mathbf{E}_f+\mathbf{E}_s$, see equations (\ref{eq:efocuspsi})-(\ref{eq:coefmatrix}). Wavelength 654\,nm, numerical aperture of the objective 1.2, $\ell\leq 250$, $|m|\leq \text{min}(\ell,200)$. Scale bars correspond to 1\,\textmu m, intensity is in arbitrary units.}
    \label{fig:excitation}
\end{figure}

With these expansions of the focal electric and magnetic fields it is straightforward to calculate the interaction of the electromagnetic field with a spherical bead of radius $R$ centered at the coordinate origin ($r=0$) and having refractive index $n_i$, so that the corresponding wave vector length is $k_i=n_i k_0$. We expand both the scattered electric field $\mathbf{E}_{s}$ and and the electric field $\mathbf{E}_{i}$ inside the bead into similar series, but with coefficients $a_{\ell m}^s,\;b_{\ell m}^s$ and $a_{\ell m}^i,\;b_{\ell m}^i$, respectively:

\begin{equation}
\begin{split}
\mathbf{E}_{s}\left(\mathbf{r}\right) = \sum_{\ell=1}^\infty\sum_{m=-\ell}^\ell \left[ a_{\ell m}^{s} \mathbf{M}^{(h^2)}_{\ell m}\left(k \mathbf{r}\right) + b_{\ell m}^{s} \mathbf{N}^{(h^2)}_{\ell m}\left(k \mathbf{r}\right) \right]
\end{split}
\end{equation}

and

\begin{equation}
\begin{split}
\mathbf{E}_{i}\left(\mathbf{r}\right) = \sum_{\ell=1}^\infty\sum_{m=-\ell}^\ell \left[ a_{\ell m}^{i} \mathbf{M}^{(j)}_{\ell m}\left(k_i \mathbf{r}\right) + b_{\ell m}^{i} \mathbf{N}^{(j)}_{\ell m}\left(k_i \mathbf{r}\right) \right].
\end{split}
\end{equation}

\noindent so that the total field outside the bead is $\mathbf{E}_f + \mathbf{E}_s$ while inside it is $\mathbf{E}_i$. By matching the four boundary conditions (tangential components of both the electric and magnetic fields are continuous across the bead's interface) we find

\begin{equation}
\label{eq:coefmatch}
\begin{split}
\begin{pmatrix}
a_{\ell m}^{i} \\ b_{\ell m}^{i} \\ a_{\ell m}^{s} \\ b_{\ell m}^{s}
\end{pmatrix} = \hat{\mathbf{L}}^{-1} \cdot \begin{pmatrix}
a_{\ell m}^{f} j_{\ell}(k R) \\ b_{\ell m}^{f} [R j_{\ell}(k R)]'/R \\ a_{\ell m}^{f} k [R j_{\ell}(k R)]'/R \\ b_{\ell m}^{f} k j_{\ell}(k R)
\end{pmatrix}
\end{split}
\end{equation}

\noindent with

\begin{equation}
\label{eq:coefmatrix}
\begin{split}
\hat{\mathbf{L}} = \begin{pmatrix}
j_{\ell}(k_{i} R) & 0 & -h^1_{\ell}(k R) & 0\\ 
0 & [R j_{\ell}(k_{i} R)]'/R & 0 & -[R h_{\ell}^1(k R)]'/R\\ 
k_{i} [R j_{\ell}(k_{i} R)]'/R & 0 & -k [R h_{\ell}^1(k R)]'/R & 0 \\ 
0 & k_{i} j_{\ell}(k_{i} R) & 0 & -k h^1_{\ell}(k R)
\end{pmatrix}
\end{split}
\end{equation}

\noindent where a prime at the square brackets denotes differentiation after $R$. This matrix equation defines the unknown coefficients $a_{\ell m}^{i}$, $b_{\ell m}^{i}$, $a_{\ell m}^{s}$ and $b_{\ell m}^{s}$ and thus the electric fields $\mathbf{E}_{i}$ and $\mathbf{E}_{s}$.\\
Figure \ref{fig:excitation} shows the resulting intensity distributions for a glass bead with radius $R=3$\,\textmu m embedded in water and varying nominal focus positions. For a nominal focus position on the optical axis, $\mathbf{r}_0=(0,0,z_{\rm{foc}})$, the maximum of the distribution is located inside the bead for $z_{\rm{foc}}\leq 4.5$\,\textmu m. With increasing $z_{\rm{foc}}$, the intensity at the nominal focus position increases but a second peak just outside the bead remains, its intensity decreasing with increasing distance between nominal focus and bead. A similar behaviour can be observed when moving the nominal focus along the $x$-direction.

\subsection{Calculating the autocorrelation function}

In this subsection we will describe how to find the autocorrelation function of a fluorescence correlation measurement when exciting and detecting fluorescence through a bead with refractive index $n$ and radius $R$. We already found the excitation intensity distribution $\left\vert \mathbf{E}_f + \mathbf{E}_s \right\vert^2$ in the previous subsection. When using the same objective for focusing and for detection, and when neglecting the Stokes shift between excitation and detection wavelength, the light collection efficiency function of confocal detection through an infinitely small pinhole would be described by the same function. For a finite pinhole, one would have to integrate this function over all focus points corresponding to the area of the pinhole, which would be numerically very demanding. Thus, we will work here in the limit of an infinitely small pinhole, so that the molecule detection function (MDF) which is given by the product of excitation intensity distribution and collection efficiency function is approximated here by the square of the excitation intensity distribution, i.e. $\left\vert \mathbf{E}_f + \mathbf{E}_s \right\vert^4$. Moreover, we will consider the diffusion of fluorescent molecules which cannot penetrate into the bead (bead surface constitutes impenetrable boundary). Thus, the correct MDF, $U(\mathbf{r})$, which has to be used for the calculation of the autocorrelation function is given by  

\begin{align}
\label{eq:mdf}
U(\mathbf{r}) = \left\{\begin{matrix} 0 & \mathrm{if} & r<R \\ \left\vert \mathbf{E}_f + \mathbf{E}_s \right\vert^4 & \mathrm{if} & r \geq R \end{matrix}\right.
\end{align}

\noindent Knowing this function, the diffusion-related part of the autocorrelation function $g(t)$ is given by eq.~\eqref{eq:autocorrelation}, where we adopt an impenetrable bead surface as the boundary condition (no flux through bead's surface). To evaluate the sixfold integral in eq.~\eqref{eq:autocorrelation}, we note that the solution of the diffusion equation,

\begin{equation}
\label{eq:diffusion}
    \frac{\partial F(\mathbf{r},t)}{\partial t} = D \Delta F(\mathbf{r},t)
\end{equation}

\noindent with initial condition $F(\mathbf{r},t=0) = U(\mathbf{r})$ is given by the first integral in eq.~\eqref{eq:autocorrelation}, 

\begin{equation}
    F(\mathbf{r},t) = \int d\mathbf{r}'\ G(\mathbf{r}-\mathbf{r}',t) U(\mathbf{r}').
\end{equation}

\noindent The overlap of this solution with the PSF gives then, in a second step, the desired autocorrelation function, using a concentration value $c$ fixed to 1:

\begin{equation}
\label{eq:overlap}
    g(t) = \int d\mathbf{r}\ U(\mathbf{r}) F(\mathbf{r},t).
\end{equation}

\noindent For solving the diffusion equation \eqref{eq:diffusion}, we discretize $U(\mathbf{r})$ on a three-dimensional evenly-spaced Cartesian grid with voxel edge length $\Delta L=50$\,nm and linear extensions $(L_x,L_y,L_z)$, which is centered on the nominal focus position $\textbf{r}_0$. In our calculations, we set $L_y=6$\,\textmu m and adjust $L_x$ and $L_z$ to values $\geq 6$\,\textmu m to capture the second peak close to the sphere (cf.\ figure \ref{fig:excitation}). The discretized values of $U$ are next arranged into a column vector $\mathbf{U}$ with $N=L_x \cdot L_y \cdot L_z / \ell^3$ elements. We discretize also the Laplace operator $\Delta$ on the same grid, and arrange its elements into a corresponding square matrix $\hat{\Delta}$ with $N$ times $N$ elements. In doing that, we implement no-flux boundary conditions on the bead's surface, and zero-values boundary conditions at the other boundaries of the discretized volume. Then, for any given time $t$, the discretized solution $\mathbf{F}$ (represented by a column vector of same size as $\mathbf{U}$) is found via matrix exponentiation,

\begin{equation}
    \mathbf{F}(t) = \exp\left(t D \hat{\Delta}\right)\cdot\mathbf{U}.
\end{equation}

\noindent Numerically, this is done by using the sparse matrix algebra implemented in \emph{Matlab}. The action of the matrix exponential is computed according to the algorithm by Al-Mohy and Higham \cite{almohy2011computing}. The autocorrelation function is finally given by the discrete approximation of the overlap integral \eqref{eq:overlap} via the scalar product

\begin{equation}
    g(t) = \Delta L^3 \mathbf{U}\cdot\mathbf{F}(t).
\end{equation}
 
\section{Numerical results}
\begin{figure}
    \centering
    \includegraphics{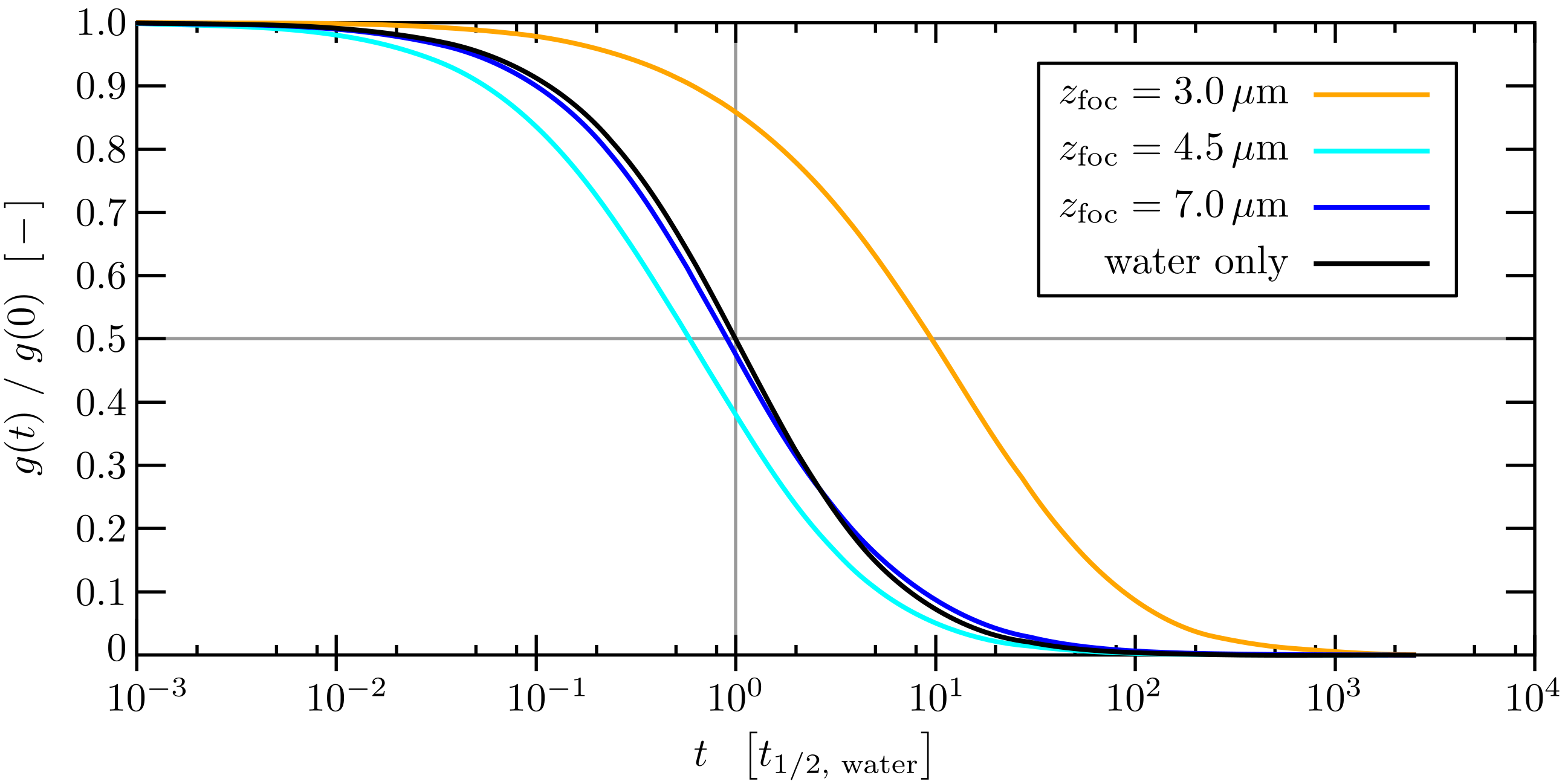}
    \caption{Normalized autocorrelation curves for a glass bead ($R=3$\,\textmu m, $n=1.52$) embedded in water ($n=1.33$) and nominal focus positions $\mathbf{r}_0=(0,0,z_{\text{foc}})$ on the optical axis. The time axis is in units of the half time of the autocorrelation curve calculated in the absence of the bead. Note that the nominal focus positions are the same as in the top row of Fig.\ \ref{fig:excitation}.}
    \label{fig:autocorrcurves}
\end{figure}

We calculate autocorrelation functions $g(t)$ as described in the previous section for glass beads (refractive index $n=1.52$) embedded in water ($n=1.33$), a wavelength of $\lambda=654$~nm, and a numerical aperture of the objective of 1.2. If one uses $-\ell \leq m \leq \ell$ in the VSH-expansion, both the computation time and the required memory increase approximately quadratically with $\ell$. However, large orders $m$ are only needed for focus positions far away from the optical axis, while large $\ell$ are needed for increasing distance between the focus position and the coordinate origin. We found a good compromise between accuracy, speed and memory usage by setting $\ell \leq 150$ and $|m| \leq 100$ for all focus positions except those with $x_\text{foc}\geq 9$~\textmu m or $x_\text{foc} > 5$~\textmu m and $z_\text{foc} \geq 7$~\textmu m, for which we set $\ell \leq 250$ and $|m| \leq 200$.

\begin{figure}
    \centering
    \includegraphics{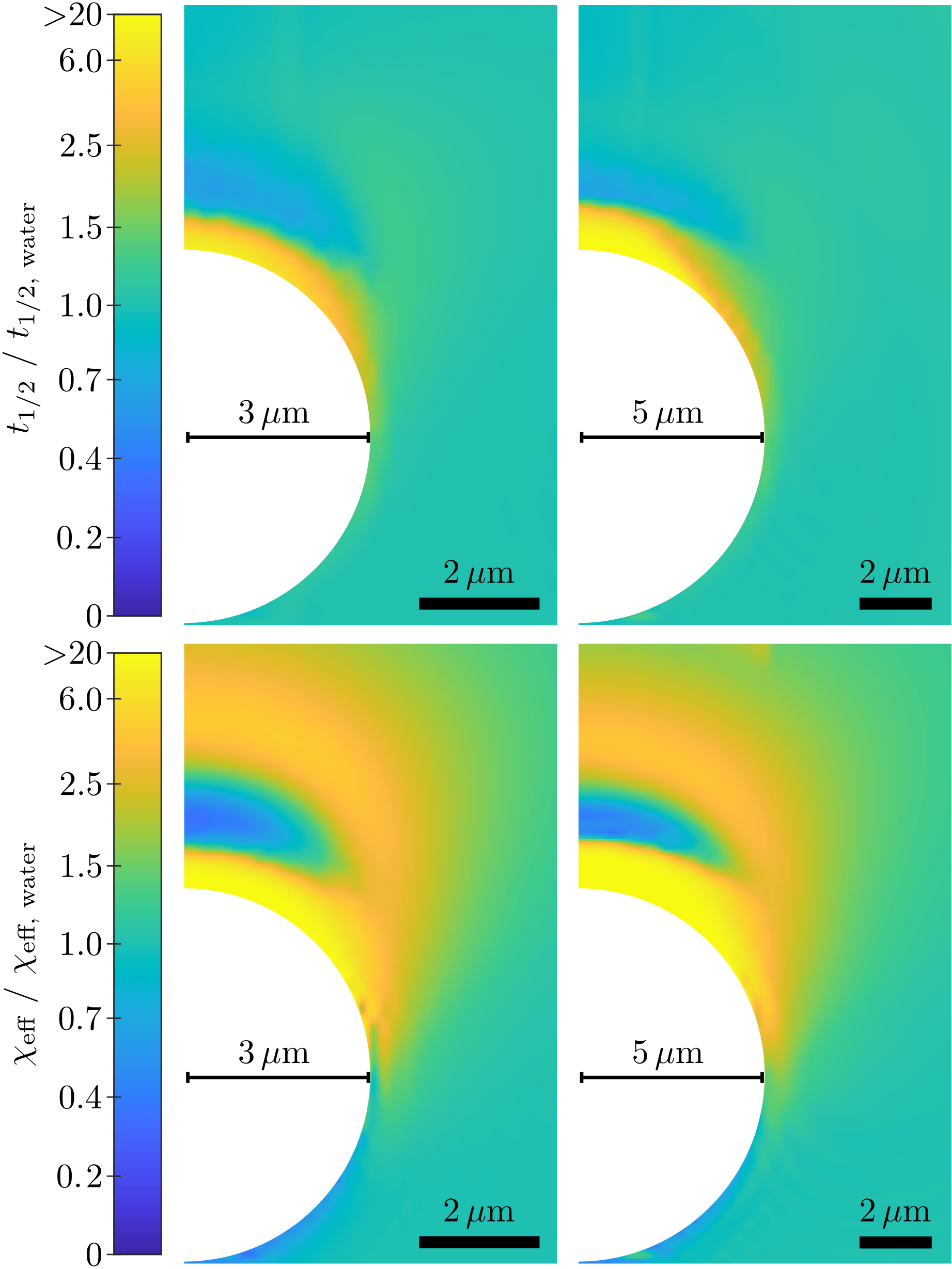}
    \caption{Maps of the halftime $t_{1/2}$ and the effective volume $V_{\text{eff}}$ calculated for glass beads ($n=1.52$) with radii $R=3$~\textmu m (left column) or $R=5$~\textmu m (right column) embedded in water. Each pixel corresponds to a different nominal focus position $\mathbf{r}_0 = (x_{\text{foc}},0,z_{\text{foc}})$ with a spacing of 40\,nm in the left column and 66.7\,nm in the right column. For nominal focus positions within the glass beads, no autocorrelation curves were calculated (white pixels). Both quantities are normalized to their values in the absence of the bead ($t_{1/2\text{, water}}$ and $\chi_{\text{eff, water}}$). The color ranges of the false-color plots are linear in the arctangent of the depicted values, which allows a better visualization of values close to one.  Note that the diffusion time is much less sensitive to the size of the confocal volume than the average number of molecules in the detection volume (which is equal to the inverse amplitude of g(t) at time zero). These calculation confirm previous experimental work where it was found that an \emph{enlargement} (not a reduction) of the effective detection volume is observed at a distance of about twice the bead diameter (distance taken from the bottom of the bead) when focusing light with an objective of numerical aperture 1.2, see Ref.~\cite{sarkar2019confocal}.}
    \label{fig:HTandVolume}
\end{figure}

Figure \ref{fig:autocorrcurves} depicts normalized curves $g(t)/g(0)$ for a bead radius of $R=3$~\textmu m and different nominal focus positions on the optical axis. For comparison, the curve calculated in the absence of the bead is also shown (black line). Both the general shape and the position of the autocorrelation functions depend on the focus position. We capture this behaviour by determining two characteristic quantities, the half time $t_{1/2}$ where the curve has decayed to half its value at $t=0$ and the effective volume $\chi_{\text{eff}}$ which is defined as

\begin{align}
    \chi_{\text{eff}} := 
        \frac{\left[\int U(\mathbf{r}) d\mathbf{r}\right]^2}{\int U^2(\mathbf{r}) d\mathbf{r}},
\end{align}

\noindent where the integrations extend over the same grid as used in the previous section. Note that we introduced the half-time of the autocorrelation decay because the lateral and axial diffusion times (usually referred to in FCS) are ill-defined. This is due to the light scattering by the bead that biases the MDF U(r), compared to its ideal shape. It is also worth to notice that $V_{\text{eff}}$ is nothing but the inverse of the amplitude of the autocorrelation curve. Figure \ref{fig:HTandVolume} shows plots of these two quantities for two different bead radii, $R=3$~\textmu m and $R=5$~\textmu m, and various nominal focus positions.

\section{Discussion and Conclusion}

We have presented a rigorous theoretical framework (based on Mie scattering theory) for the calculation of the light intensity distribution of a laser focused though a dielectric bead, and calculated autocorrelation curves for molecules diffusing through such a focus. We find an extreme sensitivity of both diffusion time and effective detection volume as a function of relative position of nominal laser focus position and bead. As can be seen in Fig.\ref{fig:HTandVolume}, both values can vary by an order of magnitude when moving the nominal focus position by only a \textmu m. However, already at short distances from the bead, the half-time approaches its ideal value for a homogeneous space without any bead. The effective volume is mostly biased at a typical distance of about one bead radius, which is consistent with previous experimental findings \cite{sarkar2019confocal}. Our result is important for two applications of FCS. In the literature, it has been proposed to use focusing through beads for reducing the effective detection volume of FCS and thus being able to record autocorrelation curves at elevated concentrations, which would be rather difficult otherwise \cite{wenger2008disposable,devilez2009three,gerard2009efficient}. Our results show that such an experiment will be extremely sensitive to the exact relative positions of nominal laser focus and bead. The other application is concerned with FCS measurements in cells and tissues, see e.g. Refs.~\cite{kim2007fluorescence,dolega2017cell}. There, one encounters optically dense obstacles such as organelles or whole cells themselves. Again, our results show that their presence may influence FCS measurements, but only very close to the obstacle, in agreement with previous experimental results \cite{sarkar2019confocal}. Thus, in such experiments, if one probes several positions across a cell or issue, it can be hoped that aberration-related variations of diffusion time or average number of emitters per volume do average out. 
\section*{Acknowledgment}

This work was performed within the project ``MICROSCATTAB'' which is jointly financed by a grant of the German Research Foundation (DFG) and the Agence National de Recherche (ANR). JR and DR acknowledge financial support by the Deutsche Forschungsgemeinschaft (DFG) through the Collaborative Research Center SFB 937 ``Collective behavior of soft and biological matter'', projects A11 and A14. JE acknowledges also funding by the DFG under Germany's Excellence Strategy - EXC 2067/1 - 390729940.

\end{document}